\documentclass[aps,prl,twocolumn,groupedaddress,showpacs]{revtex4}
\usepackage{amsmath}
\usepackage{amsfonts}
\usepackage{amssymb}
\usepackage{graphicx}

\usepackage{epsfig}%
\begin{document}
\title{Analysis of the Interface in a Nonequilibrium Two-Temperature Ising Model}
\author{P.I. Hurtado}
\affiliation{Department of Physics, Boston University, Boston, MA 02215, USA, and
Institute \textit{Carlos I} for Theoretical and Computational Physics, Universidad de
Granada, E-18071-Granada, Espa\~{n}a.}
\author{P.L. Garrido}
\author{J. Marro}
\affiliation{Institute \textit{Carlos I} for Theoretical and Computational Physics, and
Departamento de Electromagnetismo y F\'{\i}sica de la Materia, Universidad de
Granada, E-18071-Granada, Espa\~{n}a.}

\begin{abstract}
The influence of nonequilibrium bulk conditions on the properties of the
interfaces exhibited by a kinetic Ising--like model system with nonequilibrium
steady states is studied. The system is maintained out of equilibrium by
perturbing the familiar spin--flip dynamics at temperature $T$ with
completely--random flips; one may interpret these as ideally simulating some
(dynamic) impurities. We find evidence that,
in the present case, the nonequilibrium mechanism adds to the basic thermal
one resulting on a renormalization of microscopic parameters such as the
probability of interfacial broken bonds. On this assumption, we develop
theory for the nonequilibrium \textquotedblleft surface
tension\textquotedblright, which happens to show a non--monotonous behavior
with a maximum at some finite $T$. The phase diagram, as derived from this 
effective interfacial free energy, exhibits reentrant behavior. In addition,
interface fluctuations differ qualitatively from the
equilibrium case, e.g., the interface remains rough at zero--$T$, in full agreement 
with Monte Carlo simulations. We discuss
on some consequences of these facts for nucleation theory, and make some
explicit predictions concerning the nonequilibrium droplet structure.
\end{abstract}

\maketitle

\section{I. Introduction}

Interfaces that separate different homogeneous media are familiar from many
natural phenomena such as phase segregation, wetting processes, fluid
dynamics, crystal growth, and molecular beam epitaxy, for instance. In
practice, the interface may determine the system morphology and its critical
properties, or the details of time evolution, which has motivated many
specific studies during the last two
decades.\cite{Jasnow,wetting2,fluids,Bray,BarabasiStanley,crystal,wetting} The
phenomena most deeply studied so far concern interfaces that separate
\emph{equilibrium} phases. However, the actual systems of interest are seldom
at equilibrium and one often needs to be concerned with \textit{nonequilibrium
interfaces}, i.e., interfaces that separate nonequilibrium
phases.\cite{DLGinterface,Derrida,MarroDickman,tesis} This paper aims towards a
better understanding of how nonequilibrium conditions influence the properties
of an interface. The results here will be applied in a forthcoming paper to
analyze the exit from a (nonequilibrium) metastable state, which is an
interface--controlled process.\cite{next}

Mathematical complexity often compels one to deal with the simplest model
situations. In this paper, we study interfaces in a two--dimensional kinetic
Ising model. The nonequilibrium condition is obtained by perturbing the
underlying stochastic dynamics in such a way that, in general, a
nonequilibrium steady state is reached asymptotically (instead of the more
familiar thermodynamic equilibrium state).\cite{MarroDickman} We develop for
this case a simple approximation which predicts both microscopic and
macroscopic properties for the (nonequilibrium) interface, namely, the profile
or single--step height probability distribution and a (nonequilibrium)
\textquotedblleft surface tension\textquotedblright, $\sigma_{\text{ne}}.$ The
latter turns out to qualitatively differ from the equilibrium surface tension,
$\sigma_{\text{e}}.$ In particular, $\sigma_{\text{ne}}$ behaves
non--monotonously with decreasing temperature, and exhibits a maximum at some
finite temperature, unlike $\sigma_{\text{e}}$ that monotonously grows as one
cools the system. In addition, the phase diagram of the model, as derived from this effective 
surface tension, exhibits reentrant behavior. We also predict that, due to the nonequilibrium
perturbation, the interface in this model remains rough at zero temperature.
In order to understand this behavior, we analyze the shape of a droplet of the
minority phase, and conclude how the nonequilibrium condition substantially
influences the low--temperature droplet morphology. Our predictions are
compared with the results from computer simulations, which seem to firmly
support them in general.

The paper is organized as follows. The model is defined in \S \ II, and
\S \ III describes our approximation. The main results are in \S \ IV which
also contains some details of the computer simulations and a comparison of the
numerical results with theory. \S \ V is devoted to conclusions.

\section{II. The Model}

Let the square lattice $\Lambda(L_{x},L_{y})\in\mathbb{Z}^{2}$ of size
$N=L_{x}\times L_{y}$ with a binary spin variable, $s_{i}=\pm1,$ at each node
$i\in\lbrack1,N].$ We remind that the two states of $s_{i}$ may be interpreted
as corresponding to the presence or absence, respectively, of a particle at
$i;$ this happens to provide a more intuitive picture concerning the phenomena
of interest here. There is interaction between nearest--neighbor
\textit{spins} given by the Ising Hamiltonian,%
\begin{equation}
\mathcal{H}=-\sum_{\langle i,j\rangle}s_{i}s_{j}, \label{Hising}%
\end{equation}
and stochastic dynamics by single--spin flips. The latter occur with
transition rate (per unit time) given by
\begin{equation}
\omega(s_{i}\rightarrow-s_{i})=p+(1-p)\Psi\left(  \beta\Delta\mathcal{H}%
_{i}\right)  . \label{rate}%
\end{equation}
Here, $\beta=1/T$ ---we take both the coupling constant and the Boltzmann
constant equal to unity---, and $\Delta\mathcal{H}_{i}=4s_{i}(n_{i}-2)$, where
$n_{i}\in\lbrack0,4]$ is the number of up nearest--neighbor spins surrounding
$s_{i},$ i.e., $\Delta\mathcal{H}_{i}$ measures the energy cost of flipping at
$i.$ The undetermined function in (\ref{rate}) is either $\Psi\left(
\Gamma\right)  =$ e$^{-\Gamma}\left(  1+\text{e}^{-\Gamma}\right)  ^{-1}$ or
$\Psi\left(  \Gamma\right)  =\min\left[  1,\text{e}^{-\Gamma}\right]  ;$ as
indicated below, we are mostly concerned here with the first choice, except
when the second one allows for an explicit or a simpler description. In any
case, both choices lead to the results in this paper.

One may interpret that the parameter $p$ in (\ref{rate}) balances the
competition between two thermal baths: one is at temperature $T,$ while the
other induces completely random transitions as if it was at \textit{infinite
temperature}. For $p=0$, $\omega(s_{i}\rightarrow-s_{i})$ satisfies detailed
balance, and the system goes asymptotically to the equilibrium state for
temperature $T$ and energy $\mathcal{H}$. The system exhibits in this case the
familiar, Onsager critical point at $T=T_{C}(p=0)=T_{\text{O}}=2/\ln\left(
1+\sqrt{2}\right)  .$ Otherwise, $0<p<1$, the competition in (\ref{rate})
impedes canonical equilibrium and, in general, a nonequilibrium steady state
sets in asymptotically with time. A critical point is still observed in this
case, but at $T=T_{C}(p)<T_{\text{O}},$ as far as $p<p_{c}$. For
$p>p_{c}$ the system remains in the disordered phase at any $T$.
As shown below, our theoretical approach in this paper predicts 
$p_c=(\sqrt{2}-1)^2 \approx 0.1716$, in perfect agreement with previous Monte 
Carlo estimations.\cite{MarroDickman} In addition, when subject to an external 
magnetic field, this model exhibits metastable states whose strength decreases with 
increasing field, eventually becoming unstable. The spinodal field characterizing the  
limit of metastability undergoes an interesting reentrant phenomenon
as a consequence of the non-linear interplay between $T$ and $p$.\cite{PabloPRE}
We will show below that a similar reentrant behavior is observed for the phase diagram
$T_C(p)$.

In the ordered phase below $T_{C}(p),$ this system exhibits an interface for
appropriate boundary conditions. Consider, for instance, periodic boundary
conditions along the ${\hat{x}}$--direction, and open boundaries along the
${\hat{y}}$--direction. Let then the spins in the bottom (top) row to be
\emph{frozen} in the up(down)--state, while the rest of spins are allowed to
change stochastically according to (\ref{rate}). Under these conditions, an
interface eventually develops along the ${\hat{x}}$--direction that separates
up-- from down--spin rich regions located at the bottom and top of the system,
respectively. For $p=0$ (the equilibrium case), the macroscopic properties of
this interface are well--known. In particular, its scaling behavior is
characteristic of the Kardar--Parisi--Zhang universality class,\cite{KPZ} and
one knows the total free energy per unit length or surface tension.\cite{Zia}
The question is how these properties are affected by the nonequilibrium
perturbation parametrized by $p.$

\section{III. Solid-On-Solid Approximation}

\begin{figure}[t]
\centerline{
\psfig{file=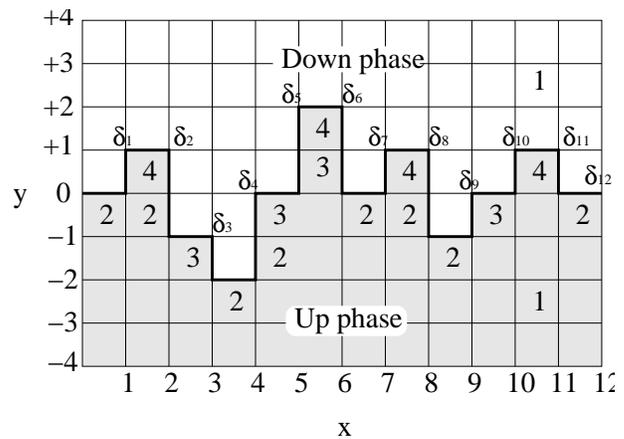,width=8cm}}\caption{{\small Example of
interface for $L_{x}=12$ with steps $\mathbf{\delta}%
=(1,-2,-1,2,2,-2,1,-2,1,1,-1,0)$. The numbers shown in the squares indicate
the class of the corresponding spin as defined in the main text. Notice that
interfacial spins can only belong to classes 2, 3 and 4.}}%
\label{sketchinterfase}%
\end{figure}

The interface is first analyzed here by adapting to our case the
\textit{solid--on--solid} (SOS) picture introduced by Burton, Cabrera and
Frank.\cite{SOS} This assumes that the interface can be described by a
single--valued discrete function completely defined by the set of interface
steps, $\{\delta_{x},x\in\lbrack1,L_{x}]\},$ as illustrated in Fig.
\ref{sketchinterfase}. No overhangs are allowed in this approximation.
Furthermore, the heights of the individual steps are assumed to be
independent. The probability of a step of height $\delta$ is assumed to be
given by%
\begin{equation}
P(\delta)=\frac{1}{z[T,p,\gamma(\phi)]}X(T,p)^{|\delta|}\text{e}^{\gamma
(\phi)\delta}.\label{pdelta}%
\end{equation}
Here $X(T,p)$ is the statistical weight associated to a broken bond in the
interface, and $\gamma(\phi)$ is a Lagrange multiplier intended to keep the
average step value at a $x$--independent value, $\langle\delta_{x}\rangle
=\tan\phi$, where $\phi$ is the \emph{average} angle between the interface and
the ${\hat{x}}$ axis.\cite{Sole,Rikvold} The function $z[T,p,\gamma]$ may be
obtained from the normalization of (\ref{pdelta}), and the \textquotedblleft
partition function\textquotedblright\ for the SOS interface follows as
$Z(T,p,\gamma)=[Xz(T,p,\gamma)]^{L_{x}}.$ This allows one to define a
\textquotedblleft thermodynamic\textquotedblright\ potential $\varphi
(T,p,\gamma)\equiv\lim_{L_{x}\rightarrow\infty}-(\beta L_{x})^{-1}\ln Z$. Our
interest is then on the \textquotedblleft free energy\textquotedblright%
\ $\sigma^{\prime}(T,p,\phi)$ ---defined as conjugated to the potential
$\varphi(T,p,\gamma)$ via a Legendre transform, which involves the variables
$\tan\phi$ and $T\gamma$--- and, in particular, on its projection $\sigma$
along the $\hat{x}$ axis. 
The nature of the above approximation is discussed
below; we now remark that our use of equilibrium words
here is only for simplicity and comfort. 
The general result is
\begin{equation}
\sigma(\phi)=T\,|\cos\phi|\Big\{\gamma(\phi)\tan\phi-\ln\frac{X(1-X^{2}%
)}{1+X^{2}-2X\cosh\gamma(\phi)}\Big\}.
\label{sigma2}
\end{equation}
In the equilibrium limit $p=0,$ one has the weight $X(T,p=0)=\text{e}%
^{-2\beta}$, and $\sigma_{\text{e}}\equiv\sigma(T,p=0,\phi)$ is the SOS
surface tension associated to the equilibrium Ising
interface.\cite{SOS,Sole,Rikvold} We shall assume in this paper that all the
main effects of the nonequilibrium perturbation $(p)$ on the interface can be
taken into account after a proper generalization of the microscopic parameter
$X(T,p)$. 
The function $\sigma_{\text{ne}}\equiv\sigma(T,p>0,\phi)$ which results
after using this generalization in (\ref{sigma2}) is therefore assumed to be the
nonequilibrium \textquotedblleft surface tension\textquotedblright\ in SOS
approximation, and this is expected to capture the macroscopic properties of
the nonequilibrium interface. The resulting $\sigma_{\text{ne}}$ is to be
interpreted as an \textit{effective}\ free--energy per unit length in
the present case that lacks of a proper bulk free--energy function.
This definition of nonequilibrium \textquotedblleft surface tension\textquotedblright\
is based on the assumption that the normalization $Z(T,p,\gamma)$ of the probability
measure associated to interface configurations in SOS picture is some sort of nonequilibrium
analog of the partition function. Similar hypothesis have been shown to yield 
excellent results when applied to other nonequilibrium models.\cite{Evans} For instance, 
in the one-dimensional asymmetric simple exclusion process (ASEP) with open boundaries,
the distribution of (complex) zeros of the steady--state normalization factor 
has been shown to obey the Lee-Yang picture of phase transitions.\cite{Evans}

In the simplest scenario ---which is consistent with the equilibrium limit---,
the weight $X(T,p)$ will not depend on $\gamma(\phi)$. An explicit relation
between $\gamma(\phi)$ and $\tan\phi$ can then be obtained. In particular,
using $\langle\delta_{x}\rangle=\tan\phi$, one finds that%
\begin{align}
\text{e}^{\pm\gamma(\phi)} &  =\frac{(1+X^{2})\tan\phi\pm S(\phi)}{2X(\tan
\phi\pm1)},\label{gamma}\\
z(\phi) &  =\frac{(1-X^{2})(1-\tan^{2}\phi)}{1+X^{2}-S(\phi)},\label{z}%
\end{align}
where $S(\phi)=[(1-X^{2})^{2}\tan^{2}\phi+4X^{2}]^{1/2}$, and we dropped an
obvious dependence on $T$ and $p$. The nonequilibrium surface tension
(\ref{sigma2}) may be now explicitly written as
\begin{align}
\sigma(\phi) &  =T\,|\cos\phi|\left\{  \left(  \tan\phi\right)  \ln
\frac{(1+X^{2})\tan\phi+S(\phi)}{2X(\tan\phi+1)}\right.  \nonumber\\
&  -\left.  \ln\frac{X(1-X^{2})(1-\tan^{2}\phi)}{1+X^{2}-S(\phi)}\right\}
.\label{sigma}%
\end{align}
Equations (\ref{pdelta}) and (\ref{sigma}) are two important properties of the
interface at the microscopic and macroscopic levels of description,
respectively. Interesting enough, (\ref{sigma}) reduces in equilibrium $(p=0)$
to the known \emph{exact} result for $\phi=0.$ It also yields a very good
approximation for any angle $|\phi|<\pi/4;$\cite{Zia,Temperley} for
$|\phi|>\pi/4,$ it is convenient to turn to the ${\hat{y}}$ (instead of
${\hat{x})}$ axis as the reference frame. $\sigma(T,p=0,\phi=0)$ is a
\textit{monotonously} decreasing function of $T$, which converges toward $2$
as $T\rightarrow0$, and vanishes at the \textit{exact} Onsager critical
temperature.
In addition, the angular dependence of $\sigma(T,p=0,\phi)$ may be used to
determine the equilibrium crystal shape via minimization of the total surface
tension for a fixed volume in a homogeneous droplet (Wulff construction).


\subsection{Statistical Weight of a Broken Bond}

At equilibrium, the weight equals the Boltzmann factor, $X(T,p=0)=\text{e}%
^{-2\beta}$ ($2$ is the energy cost of a broken bond). More generally,
$0<p<1,$ the weight of a broken bond depends on the local order surrounding
the spin at the end of the bond. One may say that $X(T,p)$ depends on the
\emph{class} of this spin. The spin $s_{i}$ is said to be of class $\eta
(s_{i},n_{i})=3-s_{i}(n_{i}-2)$, where $n_{i}$ is the number of up
nearest--neighbor spins of $s_{i}$. That is, our model may exhibit up to five
different classes of spins, $\eta=1,\ldots,5.$\cite{bordes} All the spins in a
given class are characterized by the same value of $\Delta\mathcal{H}_{\eta
}=4s_{i}(n_{i}-2)$ and, consequently, by the same rate (\ref{rate}).

The function $X_{\eta}(T,p)$ for class $\eta$ now follows straightforwardly.
Consider (\ref{rate}) with $\Psi\left(  \Gamma\right)  =\min\left[
1,\text{e}^{-\Gamma}\right]  $ which allows for a simpler and more explicit
discussion.\cite{note} If, for instance, a spin in class $\eta=1$ is flipped,
four new broken bonds appear. Since $\Delta\mathcal{H}_{1}=8$, one immediately
has that $X_{1}(T,p)=[p+(1-p)\text{e}^{-8\beta }]^{1/4}$ for the first class.
Equivalently, $X_{2}(T,p)=[p+(1-p)\text{e}^{-4\beta }]^{1/2}$ for the second
class, and the rest are characterized by the same weight as in equilibrium,
$X_{3}(T,p)=X_{4}(T,p)=X_{5}(T,p)=X(T,p=0).$ It follows that $X_{1}%
<X_{2}<X_{3,4,5}$ for any $0<p<1,$ and one may also see that $X_{\eta}\left(
T,p\rightarrow0\right)  \rightarrow\text{e}^{-2\beta},$ independent of $\eta,$
as expected.

As illustrated in Fig. \ref{sketchinterfase}, interfacial spins may only
belong to classes $2$, $3$ and $4;$ the class $1$ corresponds to spins in the
bulk of (either up or down) homogeneous regions, and the class $5$ corresponds
to typical isolated fluctuations in the bulk. Consequently, for the case in
consideration,\cite{note} only two weights, $X_{2}$ and $X_{3}=X_{4}$ are
relevant. An even simpler description ensues assuming, which amounts a
reasonable mean--field approximation, that interfacial broken bonds have an
unique statistical weight equal to the \emph{weighted} average of $X_{2}$ and
$X_{3},$ namely,%
\begin{align}
X(T,p)  &  =\Pi_{2}(T,p)X_{2}(T,p)\nonumber\\
&  +\left[  \Pi_{3}(T,p)+\Pi_{4}(T,p)\right]  X_{3}(T,p). \label{pesomedio}%
\end{align}
Here, $\Pi_{\eta}(T,p)$ is the probability of an interfacial broken bond associated 
to a spin of class $\eta.$ Alternatively, given that any bond can be arbitrarily
associated to any of the two spins at the ends, we may interpret $\Pi_{\eta
}(T,p)$ as the probability of an interfacial \emph{up} spin of class $\eta.$
Therefore, $\Pi_{2}(T,p)+\Pi_{3}(T,p)+\Pi_{4}(T,p)=1.$


\subsection{Population of Interfacial Spin Classes}

\begin{table}[t]
\centerline{
\begin{tabular}{|c|c|c|}
\hline
Step variables & Configuration & $P(\delta,\epsilon)\times {\cal Q}$   \\
\hline \hline
$\delta > 0$, $\epsilon > 0$ & \psfig{file=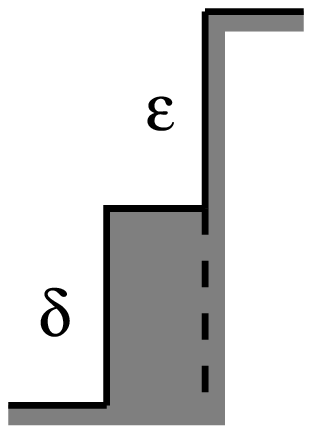,width=1cm}&
$\Lambda^{\delta+\epsilon} X_2^{\delta+\epsilon-2}X_3^3$  \\
\hline
$\delta > 0$, $\epsilon = 0$ & \psfig{file=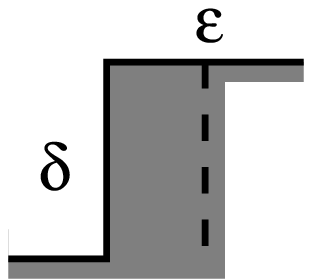,width=1cm}&
$\Lambda^{\delta}X_2^{\delta-1}X_3^2$  \\
\hline
$\delta > 0$, $\epsilon < 0$ & \psfig{file=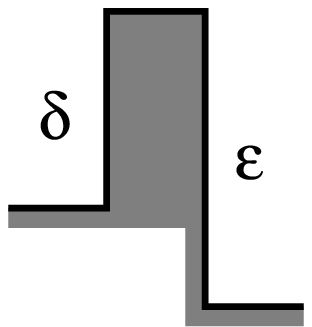,width=1cm}&
$\Lambda^{\delta+\epsilon} X_2^{\lambda - \alpha}X_3^{2\alpha+1}$  \\
\hline
$\delta = 0$, $\epsilon > 0$ & \psfig{file=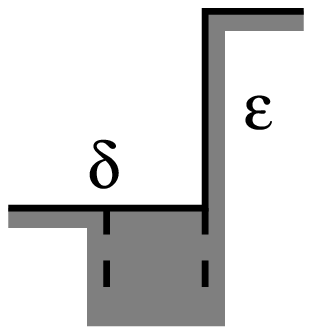,width=1cm}&
$\Lambda^{\epsilon} X_2^{\epsilon-1}X_3^2$  \\
\hline
$\delta = 0$, $\epsilon = 0$ & \psfig{file=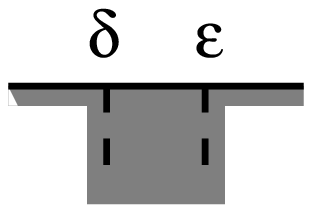,width=1cm}&
$X_2$  \\
\hline
$\delta = 0$, $\epsilon < 0$ & \psfig{file=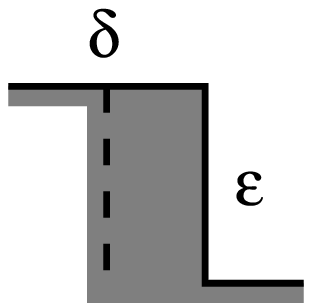,width=1cm}&
$\Lambda^{\epsilon} X_2^{|\epsilon|-1}X_3^2$  \\
\hline
$\delta < 0$, $\epsilon > 0$ & \psfig{file=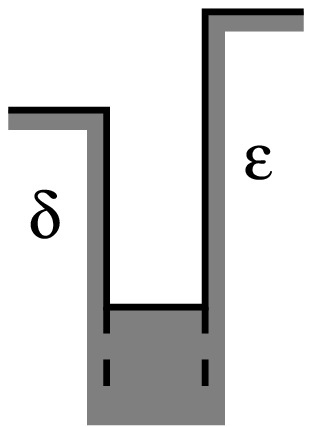,width=1cm}&
$\Lambda^{\delta+\epsilon}X_9^{\lambda - \alpha}X_3^{2\alpha+1}$  \\
\hline
$\delta < 0$, $\epsilon = 0$ & \psfig{file=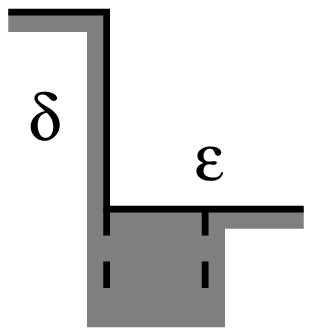,width=1cm}&
$\Lambda^{\delta} X_2^{|\delta|-1}X_3^2$  \\
\hline
$\delta < 0$, $\epsilon < 0$ & \psfig{file=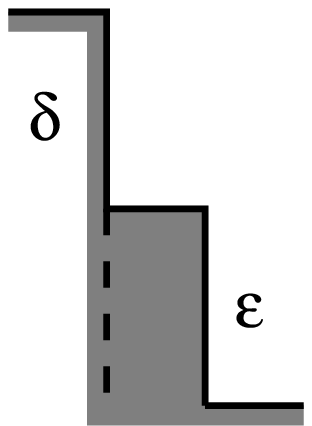,width=1cm}&
$\Lambda^{\delta+\epsilon} X_2^{|\delta|+|\epsilon|-2}X_3^3$  \\
\hline
\end{tabular}
}\caption[Different typical configurations of an interfacial spin
column.]{{\small The nine different typical configurations of an interfacial
spin column in our approximation. These configurations are defined by the
signs of the left, $\delta$, and right, $\epsilon$, steps. The last column
shows the probability $P(\delta,\epsilon)$ of each configuration. Here,
$\alpha\equiv\min(|\delta|,|\epsilon|)$ and $\lambda\equiv\max(|\delta
|,|\epsilon|);$ see the main text for other definitions..}}%
\label{configs}%
\end{table}

Next, we estimate the population densities $\Pi_{\eta}(T,p),$ which requires a
detailed counting beyond the SOS approximation. Let $P(\delta,\epsilon)$ the
joint probability that a step variable equals $\delta$ and the step variable
at its right is $\epsilon$. This completely characterizes
the population of each interfacial class in the involved column. Consider, for
instance, a case $\delta,\epsilon>0,$ as in the column between $x=4$ and $x=5$
in Fig. \ref{sketchinterfase}. This column contains $\delta+\epsilon$
interfacial spins, of which $\delta+\epsilon-2$ are of class $2$ and the other
two spins are of class $3$. On the other hand, this configuration involves
$\delta+\epsilon+1$ broken bonds, of which $\delta+1$ belong to up spins in
this column; following our convention above, the other $\epsilon$ broken bonds
may be associated to the up interfacial spins in the column between $x+1$ and
$x+2$. In order to go further in the analytical solution of the problem, let
us neglect column--column correlations by assuming that broken bonds in the
interfacial column are to be associated to interfacial spins in this column.
For $\delta,\epsilon>0$, one has $\delta+\epsilon-2$ broken bonds associated
to interfacial spins of class $2$, and three broken bonds associated to spins
of class $3$. Let us assume also that, in general, the probability of a given
interfacial column configuration is proportional to the product of
probabilities of each of the broken bonds which form it. We have that%
\[
P(\delta>0,\epsilon>0)=\frac{1}{\mathcal{Q}}\left\{  \Lambda^{\delta+\epsilon
}X_{2}^{\delta+\epsilon-2}X_{3}^{3}
\right\}  ,
\]
where $\mathcal{Q}$ is a normalization factor, and $\Lambda(\phi
)\equiv\text{e}^{\gamma(\phi)}$.\cite{OJO2} Equivalently, for $\delta>0$ and
$\epsilon<0$ one may write that%
\[
P(\delta>0,\epsilon<0)=\frac{1}{\mathcal{Q}}\left\{  \Lambda^{\delta+\epsilon
}X_{2}^{\lambda-\alpha}X_{3}^{2\alpha+1}
\right\}  ,
\]
where $\alpha=\min(|\delta|,|\epsilon|)$ and $\lambda=\max(|\delta
|,|\epsilon|)$. Table \ref{configs} shows all of the possible interfacial
column configurations, together with their statistical weights. Notice that,
as one may conclude from Table \ref{configs} and eq. (\ref{pdelta}), the
probabilities $P(\delta,\epsilon)$ converge as $p\rightarrow0$ to the SOS
equilibrium value $XP(\delta)P(\epsilon).$

\begin{table}[t]
\centerline{
\begin{tabular}{|c|c|c|c|c|}
\hline
Step variables & $n_2(\delta,\epsilon)$ & $n_3(\delta,\epsilon)$ & $n_4(\delta,\epsilon)$ & $N(\delta,\epsilon)$ \\
\hline \hline
$\delta > 0$, $\epsilon > 0$ & $\delta-1$ & 1 & 0 & $\delta$ \\
\hline
$\delta > 0$, $\epsilon = 0$ & $\delta-1$ & 1 & 0 & $\delta$ \\
\hline
$\delta > 0$, $\epsilon < 0$ & $\lambda - \alpha$ & $\alpha - 1$ & 1 & $\lambda$ \\
\hline
$\delta = 0$, $\epsilon > 0$ & 1 & 0 & 0 & 1 \\
\hline
$\delta = 0$, $\epsilon = 0$ & 1 & 0 & 0 & 1 \\
\hline
$\delta = 0$, $\epsilon < 0$ & $|\epsilon|-1$ & 1 & 0 & $|\epsilon|$ \\
\hline
$\delta < 0$, $\epsilon > 0$ & 1 & 0 & 0 & 1 \\
\hline
$\delta < 0$, $\epsilon = 0$ & 1 & 0 & 0 & 1 \\
\hline
$\delta < 0$, $\epsilon < 0$ & $|\epsilon|-1$ & 1 & 0 & $|\epsilon|$ \\
\hline
\end{tabular}
}\caption[Spin class populations for each interfacial column configuration.]%
{{\small Number $n_{\eta}(\delta,\epsilon)$ of up interfacial spins of class
$\eta=2,3,4$ for the column configuration types defined in Table
\ref{configs}. The last column shows the total number of up interfacial spins
associated to each type.}}%
\label{clasesconfig}%
\end{table}

The densities $\Pi_{\eta}(T,p)$ may be written as an average over all possible
interfacial column configurations:%
\begin{equation}
\Pi_{\eta}(T,p)=\sum_{\delta,\epsilon=-\infty}^{\infty}\pi_{\eta}%
(\delta,\epsilon)P(\delta,\epsilon). \label{Pimedio}%
\end{equation}
Here, $\pi_{\eta}(\delta,\epsilon)$ is the probability of finding an \emph{up}
spin of class $\eta\in\lbrack2,4]$ in an interfacial column characterized by
the pair $(\delta,\epsilon)$. In general, $\pi_{\eta}(\delta,\epsilon
)=n_{\eta}(\delta,\epsilon)/N(\delta,\epsilon)$, where $n_{\eta}%
(\delta,\epsilon)$ is the number of \emph{up} spins of class $\eta$ in an
interfacial column characterized by $(\delta,\epsilon)$, and $N(\delta
,\epsilon)$ is the total number of \emph{up} interfacial spins associated to
this column. Table \ref{clasesconfig} shows $n_{i}(\delta,\epsilon)$ and
$N(\delta,\epsilon)$ for all possible configurations.

The densities (\ref{Pimedio}) will depend on the average interface slope,
$\tan\phi$, through the Lagrangian multiplier $\gamma(\phi)$. This dependence
is inherited in general by the average weight of a broken bond in the
nonequilibrium regime, $X(T,p)$, see eq. (\ref{pesomedio}), and it makes the
explicit calculation of the nonequilibrium surface tension unfeasible, see
\S \ III.a. Therefore, further simplifications are needed. We shall assume
that the densities $\Pi_{\eta}(T,p)$ in eq. (\ref{Pimedio}) correspond to the
case $\tan\phi=0$, i.e., $\Lambda(\phi)\equiv\text{e}^{\gamma(\phi)}=1$. In
fact, the underlying lattice anisotropy implies that the interface tends in
general to orientate parallel to any of the principal axis, which has a lower
energy cost. Therefore, for regions of the parameter space $(T,p)$ where such
tendency is strong, it is justified to particularize the populations
$\Pi_{\eta}(T,p)$ to the case $\tan\phi=0$. On the other hand, the
parameter-space regions in which that tendency is weak are characterized by an
effective isotropy, so that particularizing to a given orientation, e.g.,
$\tan\phi=0$ is valid. This approximation amounts to assume that the relevant
orientation dependence entering the definition of the nonequilibrium surface
tension, eq. (\ref{sigma2}), comes from the dependence on $\Lambda(\phi)$ that
appears in the probability $P(\delta)$ of a step of size $\delta$ in the
interface, see eq. (\ref{pdelta}). A higher order, iterative procedure to take
into account the $\phi$-dependence of $X(T,p>0)$ would consist in: $(i)$
calculate $X^{(0)}\equiv X(T,p)$ using the $\tan\phi=0$ simplification, $(ii)$
use $X^{(0)}$ to compute $\Lambda^{(0)}(\phi)$ as a function of $\phi$ from
eq. (\ref{gamma}), $(iii)$ replace $\Lambda^{(0)}(\phi)$ in the general,
$\Lambda(\phi)$-dependent expression for $X(T,p)$, and use the so-defined
$X^{(1)}$ in (\ref{sigma}) to compute an improved approximation to the
nonequilibrium surface tension.
We do not expect this complex procedure to produce a significative improvement
of our approximation (while the resulting formulas are much more involved).

Assuming $\Lambda(\phi)=1$, and using $\pi_{\eta}(\delta,\epsilon)$ as given
in Table \ref{clasesconfig}, we obtain from eq.(\ref{Pimedio}) that%
\begin{gather}
\Pi_{2}=\frac{X_{3}^{2}}{\mathcal{Q}}\left\{  \frac{2X_{3}}{(1-X_{2})^{2}%
}+\frac{2}{1-X_{2}}\left[  2+\frac{X_{3}}{X_{2}}\ln\left(  1-X_{2}\right)
\right]  \right. \nonumber\\
+\frac{2}{X_{2}}\ln\left(  1-X_{2}\right)  +\frac{X_{2}X_{3}}{1-X_{2}}\left[
\frac{3}{X_{2}-X_{3}^{2}}+\frac{1}{1-X_{3}^{2}}\right] \label{pi2}\\
+\left.  \frac{X_{3}}{X_{2}-X_{3}^{2}}\left[  \frac{2}{X_{2}-X_{3}^{2}}%
\ln\frac{1-X_{2}}{1-X_{3}^{2}}-\frac{X_{3}^{2}}{1-X_{3}^{2}}\right]
+\frac{X_{2}}{X_{3}^{2}}\right\}  .\nonumber
\end{gather}
The same method leads to%
\begin{equation}
\Pi_{4}=\frac{X_{3}X_{2}}{\mathcal{Q}\left(  X_{2}-X_{3}^{2}\right)  }\left[
\frac{X_{3}^{2}}{X_{2}}\ln\frac{1-X_{3}^{2}}{\left(  1-X_{2}\right)  ^{2}}%
+\ln\left(  1-X_{3}^{2}\right)  \right]  , \label{pi4}%
\end{equation}
and the normalization condition gives $\Pi_{3}(T,p)=1-\Pi_{2}(T,p)-\Pi
_{4}(T,p)$. The factor $\mathcal{Q}$ follows from the normalization of
$P(\delta,\epsilon)$ for $\Lambda(\phi)=1$ as%
\begin{align}
\mathcal{Q}  &  =\frac{2X_{3}^{2}}{1-X_{2}}\left\{  2+\frac{X_{3}}{1-X_{2}%
}+\frac{X_{3}X_{2}}{X_{2}-X_{3}^{2}}\right. \nonumber\\
&  -\left.  \frac{\left(  1-X_{2}\right)  X_{3}^{3}}{(X_{2}-X_{3}^{2}%
)(1-X_{3}^{2})}+\frac{X_{3}X_{2}}{1-X_{3}^{2}}\right\}  +X_{2}.
\nonumber\label{Qnorma}%
\end{align}
In equilibrium, $p=0$, where $X_{\eta}(T,p)\rightarrow X(T,p=0)=\text{e}%
^{-2\beta}$, the above expressions for $\Pi_{\eta}(T,p)$ reduce to the known
SOS equilibrium results.\cite{Rikvold}

\section{IV. Some Results}

We summarize in this section some main results that follow from the above for
the properties of the nonequilibrium interface, and compare our predictions
with Monte Carlo simulation data.

\subsection{Microscopic Structure}

The basic SOS hypothesis is that the probability of a step of size $\delta$ in
the nonequilibrium interface is given by (\ref{pdelta}). That is, $P(\delta)$
is an exponentially--decaying function of $|\delta|$ controlled by a typical
scale which, for $\tan\phi=0,$ is ${\bar{\delta}}=\left\vert \ln X\left(
T,p\right)  \right\vert ^{-1}$ with $X$ given in (\ref{pesomedio}). In order
to check this assumption on the microscopic structure of the interface, we
performed Monte Carlo simulations of the system in \S \ II. This evolved with
time in the computer starting with two stripes of the same width of up
and down
spins, respectively, separated by a flat interface. We suppressed bulk
dynamics in these simulations, i.e., bulk, class--1 spins remained frozen to
prevent the nucleation of droplets in the bulk to interfere the interface
dynamics. This turns out to simplify notably the analysis while it does not
modify essentially the interface structure except close to the critical
temperature, where fluctuation of all sizes occur. In fact, we carefully
checked the validity of this assertion by computer simulations.

\begin{figure}[ptb]
\centerline{
\psfig{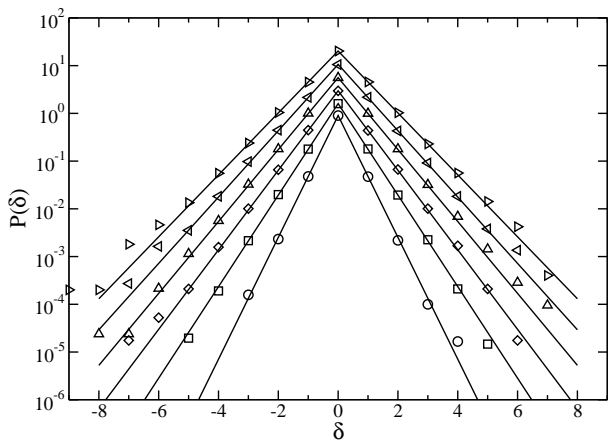}}
\caption{{\small The symbols are Monte Carlo results for the probability of step 
$\delta$ for a system of size $L_{x}\times L_{y}=256\times128$ at temperature
$T=0.3T_{\text{O}}$, with $\tan\phi=0$ and different values of the
nonequilibrium perturbation, namely, $p=0$, $0.01$, $0.02$, $0.03$, $0.04$ and
$0.05$. from bottom to top. Solid lines are the corresponding teoretical
prediction (\ref{pdelta}). For the shake of clarity, the curves are shifted by
a factor $2^{i}$, $i\in\lbrack0,5]$ in the vertical direction, where
$i=100\times p$.}}%
\label{prob-escalon}
\end{figure}

The interface was thus observed to eventually reach a steady state, in which
we measured the interface microscopic structure as a time average of
$\mathbf{\delta}=\left\{  \delta_{i},i=1,\ldots,L_{x}\right\}  $ ---taking
into account some (small) correlations observed between neighboring steps.
Fig. \ref{prob-escalon} depicts $P(\delta)$ as obtained for a (relatively
large) system at low temperature for an average interface slope such that
$\tan\phi=0$. The figure shows also our theoretical prediction, eq.
(\ref{pdelta}), revealing an excellent agreement for all values of $p.$ It is
noticeable that the typical step scale ${\bar{\delta}}$ increases as $p$
increases at fixed $T,$ i.e., the \textit{nonequilibrium noise} tends to
amplify the interface fluctuations, as one should have probably
expected.

\begin{figure}[ptb]
\centerline{
\psfig{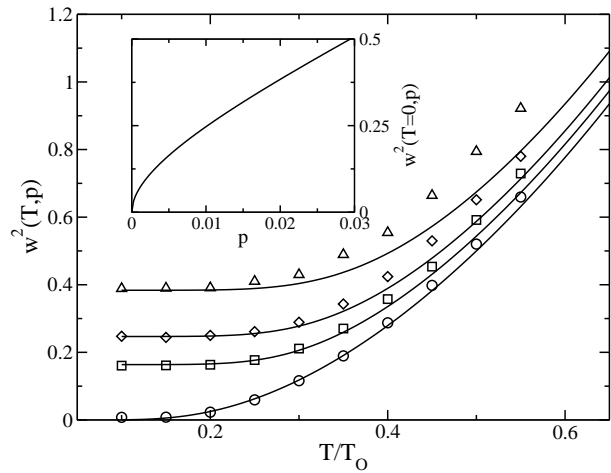}}\caption{{\small The symbols are Monte
Carlo results for the interfacial width $w^{2}(T,p)$ as a function of
temperature for a system of size $L_{x}\times L_{y}=256\times128$, $\tan
\phi=0$, and, from bottom to top, $p=0$, $0.005$, $0.01$ and $0.02$. Errorbars
are smaller that the symbol sizes. Solid lines are the corresponding
theoretical prediction. Notice the non-zero interfacial width in the low
temperature limit for the nonequilibrium system ($p>0$). The inset shows the
SOS zero-temperature limit for the interfacial width, $w^{2}(T=0,p)$, see eq.
(\ref{wT0}), as a function of $p$.}}%
\label{width}
\end{figure}

Our data is good enough to provide an estimate for the second central moment
of the step distribution, which measures the interface width, $w^{2}%
(T,p)=\langle\delta^{2}\rangle-\tan^{2}\phi$. Our theoretical prediction is%
\begin{gather}
w^{2}(T,p)=\frac{X\Lambda}{z(\phi)}\left[  \frac{1}{(1-X\Lambda)^{2}}+\frac
{1}{(\Lambda-X)^{2}}\right. \nonumber\\
+\left.  \frac{2X\Lambda}{(1-X\Lambda)^{3}}+\frac{2X}{(\Lambda-X)^{3}}\right]
-\tan^{2}\phi, \label{w2}%
\end{gather}
where $\Lambda(\phi)$, $z(\phi)$ and $X(T,p)$ are given by eqs. (\ref{gamma}),
(\ref{z}) and (\ref{pesomedio}), respectively. Fig. \ref{width} compares this
with Monte Carlo estimates. The most noticeable fact here is that $w^{2}(T,p)$
extrapolates towards a non-zero value in the low temperature limit for any
$p>0,$ contrary to the case of an equilibrium interface, which is completely
flat at zero temperature. That is, nonequilibrium fluctuations imply a rough
interface even at zero temperature. The low temperature roughness may be
estimated by realizing that, for $T\rightarrow0$ and moderate values of $p$,
$\Pi_{2}(T,p)\sim1$ from eq. (\ref{pi2}), i.e., almost all interfacial up
spins belong to class 2. Therefore, $X(T,p)\rightarrow X_{2}(T=0,p)=\sqrt{p}$
in this limit, so that one has for $\tan\phi=0$ that
\begin{equation}
w^{2}(T=0,p)\approx\frac{2\sqrt{p}}{(1-\sqrt{p})^{2}}. \label{wT0}%
\end{equation}
The inset in Fig. \ref{width} depicts this behavior which is in full agreement
with our Mote Carlo values. The only significant differences we found between
theory and simulations are at high enough temperatures, as illustrated in Fig.
\ref{width}. This is to be attributed to step--step correlations as the
critical temperature is approached.

\subsection{Macroscopic Behavior}

\begin{figure}[ptb]
\centerline{
\psfig{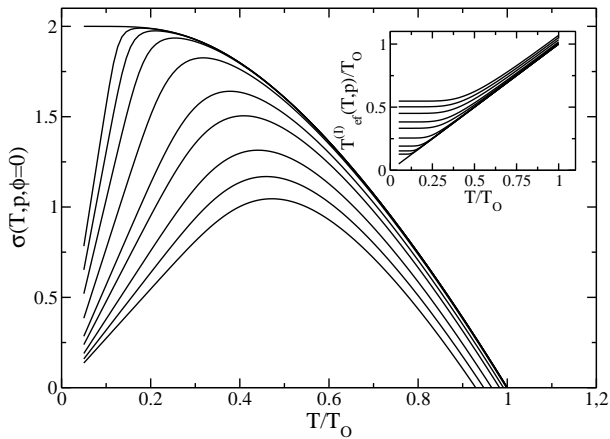}}\caption{{\small Main
graph: Theoretical prediction for the surface tension as a function of
temperature for, from top to bottom, $p=0$, $10^{-6} $, $10^{-5}$, $10^{-4}$,
$10^{-3}$, $5\times10^{-3}$, $10^{-2}$, $2\times10^{-2}$, $3\times10^{-2}$ and
$4\times10^{-2}$. Notice that, for any ---even small--- $p>0,$ the surface
tension behaves non-monotonously, contrary to the equilibrium case. Inset: The
effective interface temperature, $T_{\text{ef}}^{\text{(I)}}$, as defined in the main text,
as a function of $T$ for the same values of $p$ than in the main graph. Notice
that $T_{\text{ef}}^{\text{(I)}}(T,p>0)$ strongly deviates from $T$ in the low temperature
regime.}}%
\label{tensionsup}%
\end{figure}

A principal interface macroscopic property is the surface tension. The
prediction (\ref{sigma}) is illustrated in Fig. \ref{tensionsup} for $\phi=0.$
It is remarkable the essential difference occurring at low $T$ between the
equilibrium and nonequilibrium cases.

The surface tension $\sigma_{\text{ne}}\left(  T,p>0;\phi=0\right)  $ exhibits
non-monotonous behavior as a function of $T,$ with a maximum at a temperature
which depends on the intensity of the nonequilibrium perturbation,
$T_{max}(p).$ For $T<T_{max}(p),$ $\sigma_{\text{ne}}$ decreases as one cools
the system. This \emph{anomalous} behavior turns out to play a fundamental
role in understanding the exit from metastable states in this
system,\cite{tesis,next} and it is likely it may help the understanding of the low
temperature behavior in other complex systems concerning nucleation and growth processes.

We devised the following indirect method to check our predictions for the
nonequilibrium surface tension. The system defined in \S \ II happens to
exhibit long--lived metastable states in the ordered phase when subject to a
small negative external magnetic field.\cite{next,PabloPRE} The exit from this state is
a highly inhomogeneous process that proceeds via the nucleation and growth of
one or several droplets of the stable phase within the metastable sea. Droplet
nucleation is controlled by the competition between the surface tension, which
hinders the droplet growth, and the bulk \textquotedblleft free
energy\textquotedblright, which favours it. Consequently, small droplets
---having a large surface/volume ratio--- tend to shrink, while the larger
ones tend to grow. The critical droplet size, $R_{c}(T,p)$, separates these
two regimes. Following further the trend in equilibrium theory,\cite{next} one
may assume that an effective macroscopic potential controls the escape from
the metastable state for $0<p<1,$ and that
\begin{equation}
R_{c}(T,p)=\frac{(d-1)\,\sigma(T,p)}{2m_{s}(T,p)|h|}. \label{Rc}%
\end{equation}
Here, $d$ is the system dimensionality, $h$ is the applied magnetic field,
$\sigma(T,p)$ stands for the zero--field surface tension along one of the
lattice axis, and $m_{s}(T,p)$ is the spontaneous (positive) magnetization for
$h=0.$ The latter may be approximated by mean field--theory,\cite{PabloPRE} and the surface
tension may then be obtained from a Monte Carlo estimate of $R_{c}(T,p).$
\begin{figure}[ptb]
\centerline{
\psfig{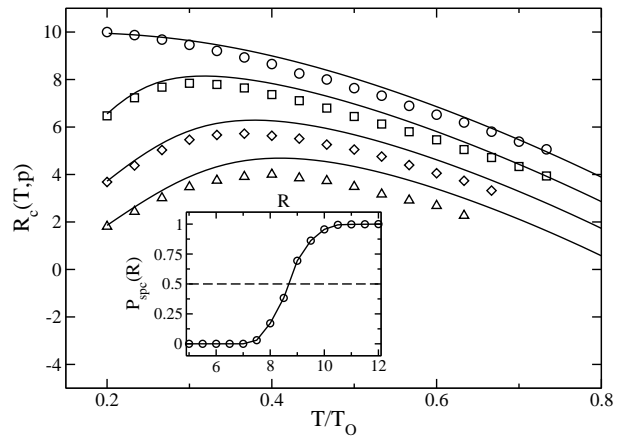}}\caption{{\small Critical
droplet size, $\mathcal{R}_{c}$, as a function of temperature for a system of
size $L=53$, with periodic boundary conditions, subject to a magnetic field
$h=-0.1,$ and, from top to bottom, $p=0$, $0.001$, $0.005$ and $0.01$. The
symbols are Monte Carlo results obtained as an average over $N_{exp}=1000$
independent \textquotedblleft experiments\textquotedblright. The solid curves
correspond to the theoretical prediction (\ref{Rc}). For the shake of clarity,
results for the $n$-th value of $p$, $n=1,\ldots,4$ (using the above indicated
order) have been shifted by $(1-n)$ units along the $\hat{y}$ axis. The inset
shows Monte Carlo results for the probability that a droplet of radius $R$ is
supercritical, $P_{spc}(R)$, as a function of $R$ for a system of size $L=53$
at $T=0.4T_{\text{O}}$, $p=0$, and $h=-0.1.$ This corresponds to 10}$^{3}%
${\small \ independent experiments for each value of $R$. In all cases, error
bars are smaller than the symbol sizes.}}%
\label{gotacrit}
\end{figure}

In order to perform this computation, consider our system in a square lattice
with periodic boundary conditions along both the ${\hat{x}}$ and ${\hat{y}}$
directions. The Hamiltonian is now $\mathcal{H}^{\prime}=\mathcal{H}-h\sum
_{i}s_{i}$, with $\mathcal{H}$ given in (\ref{Hising}) and $h<0.$ All the
spins are initially up, except for a \emph{square} droplet of down spins of
side $2R$ which represents the stable phase. This is let to evolve according
to (\ref{rate}). The state is highly unstable, so that any
\textit{subcritical} initial cluster, i.e., $R<R_{c}(T,p),$ will very quickly
shrink, while a \textit{supercritical} one, $R>R_{c}(T,p),$ will rapidly grow
to cover the whole system. Since our dynamics is stochastic, we define the
probability that a droplet of size $R$ is supercritical, $P_{spc}(R).$ This is
measured in practice by simply repeating many times the simulation and
counting the number of times that the initial droplet grows to cover the
system. The critical droplet size is defined by $P_{spc}(R_{c})=0.5.$ As
observed in the inset of Fig. \ref{gotacrit}, $P_{spc}(R)$ sharply goes from
$0$ to $1,$ which allows a relatively accurate estimate of $R_{c}.$

Fig. \ref{gotacrit} compares our Monte Carlo results for $R_{c}(T,p)$ with the
theoretical prediction from (\ref{Rc}) using $\sigma(T,p)=\sigma_{\text{ne}%
}(T,p;\phi=0)$ as given by eq. (\ref{sigma}). The agreement is rather good,
and we confirm that $R_{c}(T,p)$ behaves non-monotonously with $T$ in the
nonequilibrium regime.

\begin{figure}[ptb]
\centerline{
\psfig{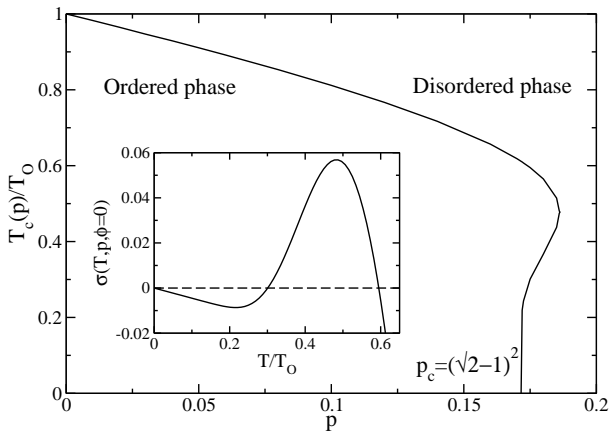}}\caption{{\small Phase diagram
of the model as obtained from the nonequilibrium surface tension. Notice the reentrant
behavior of $T_C(p)$ for $p_c<p<p_c^*$, with $p_c=(\sqrt{2}-1)^2$ and $p_c^*\approx 0.18625$.
Inset: $\sigma_{\text{ne}}(\phi=0)$ as a function of $T$ for $p=0.175>p_c$. Notice the negative
slope at low-$T$ and the intermediate temperature regime where $\sigma_{\text{ne}}$ 
is positive.}}
\label{diag}
\end{figure}

The temperature dependence of $\sigma_{\text{ne}}(T,p;\phi=0)$ may be used to 
compute the phase diagram of the model, $T_C(p)$.\cite{Hartmann} In equilibrium, the 
interface free energy approaches zero as $T\rightarrow T_O$; for 
$T>T_O$ there is no surface tension because there exist only one disordered 
phase.\cite{Hartmann} Therefore, if as assumed in this paper
$\sigma_{\text{ne}}$ captures the macroscopic properties of the nonequilibrium interface, we may
identify $T_C(p)$ as the temperature (other than $T=0$) for which 
$\sigma_{\text{ne}}(T,p,\phi=0)=0$. This is done in Fig. \ref{diag}. In particular,
we may ask about the nonequilibrium parameter $p_c$ above which no ordered phase exists at
low $T$. For $p>0$ and $T\rightarrow 0$, $\sigma_{\text{ne}}(T,p;\phi=0)\sim \alpha(p) T$ 
(see Fig. \ref{tensionsup} and eq. (\ref{sigma})). The slope 
$\alpha(p)$ decreases monotonously with $p$, and the condition $\alpha(p_c)=0$ signals
the onset of disorder at low temperature. This yields $p_c=(\sqrt{2}-1)^2\approx 0.1716$,
in excellent agreement with previous Monte Carlo simulations.\cite{MarroDickman} 
For $p>p_c$ one does not expect low-$T$ order. However, the non-monotonous temperature 
dependence of $\sigma_{\text{ne}}$ (see inset in Fig. \ref{diag}) involves the emergence 
of an intermediate-$T$ region for $p_c<p<p_c^*\approx 0.18625$ where order sets in, as opposed 
to the low-$T$ and high-$T$ disordered phases, see Fig. \ref{diag}. This reentrant behavior 
of $T_C(p)$ is similar in spirit to the one reported in Ref. \cite{PabloPRE} for the spinodal 
field characterizing the limit of metastability in this model, and it is reminiscent of the
reentrant phase diagram observed in systems subject to multiplicative noise.\cite{reentrant}

In order to gain some intuition on the physical origin of the anomalous
low--temperature behavior of the nonequilibrium surface tension, let us write
the statistical weight of an interfacial broken bond as
\begin{equation}
X(T,p)=\text{exp}\left[  -2\beta_{\text{ef}}^{\text{(I)}}\right]  , \label{teffi}%
\end{equation}
which defines an \textit{interface effective temperature}. The function
$T_{\text{ef}}^{\text{(I)}}(T,p)=1/\beta_{\text{ef}}^{\text{(I)}}$ is illustrated in the inset of Fig.
\ref{tensionsup}. One first realizes that $T_{\text{ef}}^{\text{(I)}}(T,p>0)>T$ for any
$T\in\lbrack0,T_{\text{O}}]$. That is, the nonequilibrium interface endures an
effective temperature larger than the thermodynamic one. On the other hand,
one identifies two different regimes in the inset of Fig. \ref{tensionsup} at
given $p.$ At high-$T$, where thermal fluctuations dominate over the nonequilibrium
noise, $T_{\text{ef}}^{\text{(I)}}$ is proportional to $T$. However, at low enough $T$ 
the nonequilibrium noise dominates; in this case
$T_{\text{ef}}^{\text{(I)}}$ deviates from $T$ and tends to a saturating, constant value 
which depends on $p.$
Following the method above to obtain the zero-temperature interfacial width, we conclude
that
\begin{equation}
\lim_{T\rightarrow0}T_{\text{ef}}^{\text{(I)}}(T,p)\approx-\frac{4}{\ln p}>0.
\label{teff}
\end{equation}
It is also remarkable that the onset of the deviation of $T_{\text{ef}}^{\text{(I)}}$ from
$T$ for a given $p$ coincides with the maximum observed for the nonequilibrium
surface tension; see Fig. \ref{tensionsup}. In fact, since $T_{\text{ef}}^{\text{(I)}}$ is a 
small constant for $0<T\ll T_C(p)$ and $p\ll p_c$, we may expand $\sigma_{\text{ne}}$ at low-$T$
to obtain $\sigma_{\text{ne}} \sim (T/T_{\text{ef}}^{\text{(I)}})\sigma_{\text{e}} + {\cal O}(TX)$.
On the other hand, for $0\ll T <T_C(p)$ and the same $p\ll p_c$, the quotient 
$T/T_{\text{ef}}^{\text{(I)}}\equiv 1-\alpha$ is a constant, with $0<\alpha \ll 1$. Using $\alpha$
as small parameter now, a high-$T$ expansion of $\sigma_{\text{ne}}$ yields 
$\sigma_{\text{ne}}\sim (T/T_{\text{ef}}^{\text{(I)}})\sigma_{\text{e}} + {\cal O}(\alpha)$. In both
limits the corrections to the asymptotic behavior  $\sigma_{\text{ne}}\sim (T/T_{\text{ef}}^{\text{(I)}})\sigma_{\text{e}}$
are small. Now, since $T_{\text{ef}}^{\text{(I)}} \sim \textrm{constant}$ and 
$\sigma_{\text{e}} \sim 2(1-T\textrm{e}^{-2/T})$ as $T \rightarrow 0$, one expects 
$\sigma_{\text{ne}}$ to be an increasing function of $T$ in this limit. On the other hand, 
$T/T_{\text{ef}}^{\text{(I)}} \sim \textrm{constant}$ in the high-$T$ limit, so $\sigma_{\text{ne}}$ depends
on $T$ as $\sigma_{\text{e}}$ does, i.e. $\sigma_{\text{ne}}$ decreases with $T$ for high enough $T$. Therefore
one would expect a non-monotonous $T$-dependence of $\sigma_{\text{ne}}$, with a maximum at 
$T_{max}(p)$, which roughly coincides with the crossover observed in $T_{\text{ef}}^{\text{(I)}}(T,p)$.
In this way, the anomalous $T$-dependence of $\sigma_{\text{ne}}$ can be traced back to the 
crossover between a $T$-dominated, high-$T$ regime and a $p$-dominated, low-$T$ region, 
as captured by $T_{\text{ef}}^{\text{(I)}}$.

\subsection{Droplet Shape}

The droplet shape is controlled by the need to minimize the total surface
tension at constant droplet volume. For isotropic systems, this implies
spherical shape. In our case, however, the surface tension depends on the
orientation of the interface with respect to a privileged axis, $\sigma
(\phi).$ Consequently, the shape adjusts itself to take advantage of the low
free energy cost of certain interface orientations, which produces droplets
with a crystal--like appearance which depends on temperature and other
parameters.\cite{Rottman} We apply next the Wulff construction \cite{Wulff} to
obtain information concerning the nonequilibrium droplet shape.
\begin{figure}[ptb]
\centerline{
\psfig{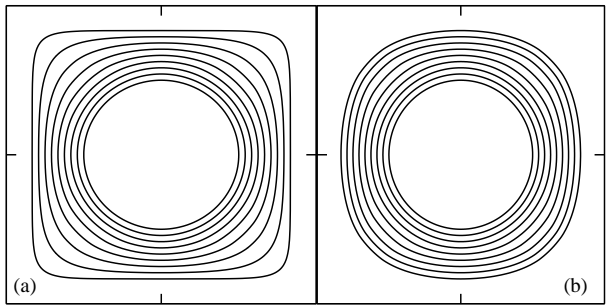}}\caption{{\small (a)
Shape of a droplet, as obtained from the Wulff construction, for $p=0$
(equilibrium) at, following to the centre, $T/T_{\text{O}}=0.1$, $0.2$, $0.3$,
$0.4$, $0.5$, $0.6$, $0.7$, $0.8$ and $0.9$. For the shake of clarity we have
rescaled the droplet according to its temperature. (b) The same as in (a), but
for the nonequilibrium model with $p=0.01$.}}%
\label{formas}%
\end{figure}

The method essentially consists in considering the polar curve $\sigma(\phi)$,
$\phi\in\lbrack0,2\pi]$, and drawing through its points a line perpendicular
to the radius. The interior envelope to these lines determines the droplet
radial function in polar coordinates, $R(\theta).$ More specifically, one may
write parametrically:\cite{D}
\begin{eqnarray}
R(\theta) & = & R_{0}[x^{2}(\phi)+y^{2}(\phi)] ^{1/2}, \\
x(\phi)  &  = & \sigma(\phi)\,\cos\phi-\frac{\text{d}\sigma(\phi)}{\text{d}%
\phi}  \sin\phi,\nonumber\\
y(\phi)  &  = & \sigma(\phi)\,\sin\phi-\frac{\text{d}\sigma(\phi)}{\text{d}%
\phi}  \cos\phi,\nonumber\\ 
\tan\theta & = & \frac{y(\phi)}{x(\phi)},\nonumber
\end{eqnarray}
where $R_{0}$ is a fixed length scale, and $\sigma(\phi)$ is the surface
tension. We approximate the latter by (\ref{sigma}). This is needed only for
the angular interval $\phi\in\lbrack0,\pi/4]$, since one may extend then to
the whole circumference by straightforward symmetry considerations. The result
is singular for angles $\phi=(2n+1)\pi/4,$ $n=0,\ldots,3,$ which gives rise to
angular intervals around $\theta=(2n+1)\pi/4$, $n=0,\ldots,3$, where
$R(\theta)$ is not defined. Therefore, one considers the analytical
continuation $r(\theta)$ such that, in particular, d$r/$d$\theta=0$ at
$\theta=\pi/4$ \ as required by symmetry. This, together with continuity and
analyticity, leads to a second order polynomial and its coefficients,
$r(\theta)=a\theta^{2}+b\theta+c.$

Fig. \ref{formas} illustrates the result. In equilibrium, $p=0,$ the droplet
tends to become squared as $T\rightarrow0$ due to the underlying lattice
anisotropy, while it recovers the (isotropic) spherical shape for
$T\gtrsim0.5T_{\text{O}}.$ In nonequilibrium, the droplet adopts a shape which
is intermediate between a circle and a square. This is again due to the fact
that the temperature $T_{\text{ef}}^{\text{(I)}}$ that the interface \emph{feels} does not
go to zero as $T\rightarrow0$ for any $p>0.$

\begin{figure}[ptb]
\centerline{
\psfig{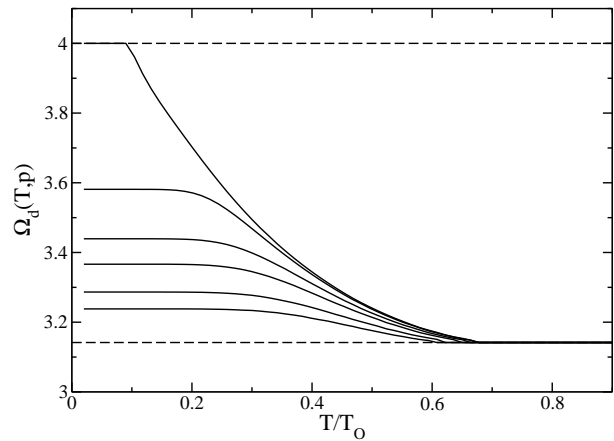}}\caption{{\small The form factor
$\Omega(T,p)$ as a function of temperature for, from top to bottom, $p=0$,
$0.001$, $0.005$, $0.01$, $0.02$ and $0.03$. The top (bottom) line corresponds
to the squared (circular) droplet.}}%
\label{fforma}%
\end{figure}

A more quantitative description is provided by the droplet form factor,
$\Omega_{d}(T,p).$ This is defined via the equality $V=\Omega_{d}%
(T,p)\mathcal{R}^{d},$ where $\mathcal{R}\equiv R(\theta=0)$ is a measure of
the droplet radius, and $V$ is the droplet volume. For a two--dimensional
system,%
\begin{equation}
\Omega(T,p)=4\int_{0}^{\pi/4}\text{d}\theta\left(  \frac{R(\theta)}%
{R(0)}\right)  ^{2}. \label{omega}%
\end{equation}
The square and circular droplets are characterized by $\Omega=4$ and $\pi,$
respectively. Fig. \ref{fforma} shows $\Omega(T,p)$ as a function of $T$ for
different values of $p.$ This clearly demonstrates that $\Omega\left(
T,p\right)  $ goes to 4 (squared shape) in the low--$T$ limit for $p=0,$ but
the tendency is towards a smaller value for any $p>0.$ That is, unlike in
equilibrium at low enough temperature, no facets are expected in a
nonequilibrium droplet.

\section{Conclusions}

This paper deals with the influence of a simple nonequilibrium condition,
which may ideally represent the situation in a class of disordered
systems,\cite{MarroDickman} on the microscopic and macroscopic properties of a
complex interface. It is assumed that the probability of a discrete change by
$\delta$ at the (nonequilibrium) interface is proportional to $X^{\left\vert
\delta\right\vert },$ where $X=X(T,p)$ is the statistical weight of a
interfacial broken bond for temperature $T$ and nonequilibrium perturbation
$p.$ This is expressed $X=\sum_{\eta}\Pi_{\eta}X_{\eta}$ in terms of the
probability $\Pi_{\eta}$ that an interfacial broken bond ends at a spin
surrounded by certain degree, $\eta,$ of local order. 

Our main hypothesis consists in assuming that the nonequilibrium system is
attempting to minimize a surface and, consequently, one may translate here the
equilibrium formalism. It is also assumed that, at least for the model studied
in this paper, the specific nonequilibrium mechanism simply adds to the basic
thermal mechanism in such a way that it may be incorporated in a
non--perturbative manner to the microscopic parameter $X(T,p).$ Therefore,
this contains \textit{all} the information concerning the effect of the
nonequilibrium perturbation $p$ on the interface, and it follows that a SOS
theory based on $X(T,p)$ yields the micro- and macroscopic behavior of the
nonequilibrium interface. In this way, one may finally obtain an explicit
expression for the relevant nonequilibrium surface tension, $\sigma
_{\text{ne}}\left(  T,p\right)  .$

Regarding the microscopic interface structure, the nonequilibrium
\textit{noise} turns out to enhance interfacial fluctuations. In particular,
the typical scale for interfacial fluctuations increases with $p.$ It is also
demonstrated that the nonequilibrium interface remains rough in the zero--$T$
limit, contrary to the equilibrium case. These are theoretical predictions in
full agreement with Monte Carlo simulations.

Regarding macroscopic behavior, $\sigma_{\text{ne}}\left(  T,p\right)  $
exhibits \emph{anomalous} behavior at low $T$ (for any $p>0).$ In particular,
$\sigma_{\text{ne}}$ is a non-monotonous function of $T$ with a maximum at
$T=T_{max}(p),$ and $\sigma_{\text{ne}}$ decreases as the system is cooled
further below $T_{max}(p).$ This counter--intuitive prediction is also
confirmed indirectly by Monte Carlo simulations. That is, we estimated
numerically the critical droplet size, $R_{c}(T,p),$ which is expected to be
proportional to the surface tension,\cite{next} as the system exits from a
metastable state. Some intuition on the origin of this anomaly is obtained by
defining an interface effective temperature which importantly deviates from
$T.$ In this way, the non-monotonous $T$-dependence of $\sigma_{\text{ne}}$ can be related to
a crossover between two different temperature regimes: a low-$T$ region dominated by the nonequilibrium noise, where $\sigma_{\text{ne}}\propto T$, and a high-$T$ regime dominated by thermal fluctuations, where $\sigma_{\text{ne}}\propto\sigma_{\text{e}}$.
The shape of the nonequilibrium droplet as obtained by a Wulff
construction also reflects the anomaly of $\sigma_{\text{ne}}.$ We find, in
particular, that droplets at very low temperature tend to minimize more their
surface under the nonequilibrium condition.

These details are essential to nucleation theory. Therefore, we expect that
the anomalous low--temperature behavior of the nonequilibrium surface tension
described above may be relevant to many physical processes such as the ones
mentioned in \S \ I. The possible utility of our results here will be
addressed in a forthcoming paper concerning the relaxation of a nonequilibrium
system from a metastable state.\cite{next}

Finally, the results in this paper are explicitely obtained for a square lattice. Some 
caution should be used before generalizing, since there are examples when the shape and 
properties of a nonequilibrium interface depend strongly on the geometry 
of the host lattice.\cite{lattice} However, we believe that the phenomenology here described should hold
for more lattice geometries other than square, provided that the (nonequilibrium) interface endures an effective temperature 
with the same qualitative properties than the one discussed above, i.e. $T_{\text{ef}}^{\text{(I)}}$ saturates to a 
constant, $p$-dependent value as $T \rightarrow 0$ and is proportional to $T$ at high enough temperature.

\section*{Acknowledgments}

We acknowledge very useful discussions with M.A. Mu\~{n}oz, and financial
support from MCYT-FEDER project BFM2001-2841. P.I.H.
also thanks support from the MECD.


\begin{thebibliography}{99}
\bibitem {note}For $\Psi\left(  \Gamma\right)  =$ e$^{-\Gamma}\left(
1+\text{e}^{-\Gamma}\right)  ^{-1},$ all the functions $X_{\eta}(T,p)$ differ
from each other, which makes the expressions and discussion concerning this
point much more involved. However, the numerical differences between the two
rates are negligible for all practical purposses here, so that we found it
convenient to explicitly illustrate in this section the case $\Psi\left(
\Gamma\right)  =\min\left[  1,\text{e}^{-\Gamma}\right]  $ only.
\bibitem {OJO2}The SOS approximation would imply that $P(\delta,\epsilon
)=XP(\delta)P(\epsilon)$ with $P(\delta)$ given in (\ref{pdelta}). One may
then estimate $\Pi_{\eta}(T,p)$ as a function of $X(T,p),$ and a trascendental
equation for $X(T,p)$ ensues which cannot be solved analytically. The higher
order approximation that we present here provides an explicit expression.
\end{thebibliography}
\end{document}